\documentclass[onecolumn,preprintnumbers,amsmath,amssymb,superscriptaddress]{revtex4}
\usepackage{graphicx}
\usepackage{dcolumn}
\usepackage{bm}
\usepackage{longtable}

\usepackage{epsfig}
\usepackage{amssymb}
\usepackage{amsmath}
\usepackage{amsthm}
\usepackage{latexsym}
\usepackage{graphicx}
\usepackage{epstopdf}
\usepackage{subfigure}
\usepackage[pdftex,hyperfigures,breaklinks,colorlinks,citecolor=blue,linkcolor=red]{hyperref} 

\def\be{\begin{equation}}
\def\ee{\end{equation}}
\def\ball{\begin{align}}
\def\eall{\end{align}}
\def\bc{\begin{center}}
\def\ec{\end{center}}

\def\<{\bigl\langle}
\def\>{\bigr\rangle}

\def\s{\sigma}

\begin{document}


\title{Dynamic message-passing approach for kinetic spin models with reversible dynamics}

\author{Gino Del Ferraro}
\affiliation{Department of Computational Biology, AlbaNova University Centre, SE-106 91 Stockholm, Sweden}
\author{Erik Aurell}
\affiliation{Department of Computational Biology, AlbaNova University Centre, SE-106 91 Stockholm, Sweden}
\affiliation{ACCESS Linnaeus Centre, KTH-Royal Institute of Technology, SE-100 44 Stockholm, Sweden}
\affiliation{Depts of Information and Computer Science and Applied Physics
and Aalto Science Institute, Aalto University, P.O. Box 15400, FI-00076 Aalto, Finland}

\date{\today}

\begin{abstract}
A method to approximately close the dynamic cavity equations for synchronous reversible dynamics on a locally tree-like topology is presented. The method builds on \textit{(a)} a graph expansion to eliminate loops from the normalizations
of each step in the dynamics, and \textit{(b)} an assumption that a set of auxilary probability distributions on 
histories of pairs of spins mainly have dependencies that are local in time. The closure is then effectuated by 
projecting these probability distributions on $n$-step Markov processes. The method is shown in detail 
on the level of ordinary Markov processes ($n=1$), and outlined for higher-order approximations ($n>1$). Numerical validations of the technique are provided for the reconstruction of the transient and equilibrium dynamics of the kinetic Ising model on a random graph with arbitrary connectivity symmetry.  
\end{abstract}

\maketitle  

\section{Introduction}
Disordered spin systems are an important class of models able to catch and reproduce a large range of phenomena from phase transitions in magnets and amorphous systems \cite{crisanti2007amorphous} to protein folding in biology \cite{goldstein1992optimal}, social media \cite{facchetti2011computing,kirkpatrick2012social}, epidemic spreading \cite{pastor2001epidemic}, immune and neural networks \cite{amit1992modeling}, applications in finance and optimisation problems \cite{coolen2005mathematical,mezard2002analytic,PhysRevLett.87.127209}. 
In the thermodynamic limit these systems have rich and fascinating repertoires of static and dynamic behaviour including
the clustering~\cite{mezard2002analytic} or shattering~\cite{achlioptas2010algorithmic} transition, ergodicity breaking and  
ageing~\cite{bouchaud1997out}. To systematically describe their static properties the replica method (for fully connected systems) and the cavity method (for dilute systems) were developed~\cite{mezard2009information}.
General techniques to systematically study the dynamics of single finite systems in this class have however
been less developed and in practice mostly limited to dynamic mean-field theories~\cite{Kappen98,kappen2000mean}, 
path integral techniques~\cite{de1978dynamics,Sompolinsky1982,sommers1987path,RoudiHertz11JSAT}, large deviation approaches \cite{del2014perturbative}  
and, above all, numerical simulation~\cite{Gillespie77}.

The cavity method \cite{mezard2009information,yedidia2003understanding} here holds
a special place as it has become the method of choice to solve the statics of models on sparse networks,
while for dynamics it was long restricted to dynamics on fully asymmetric graphs~\cite{derrida1987exactly,Mezard11,aurell2011three}.
The main problem is that while the cavity technique reduces the complexity ``in space'' 
(number of terms in an approximate computation of marginals), there remains
a complexity ``in time'' (cardinality of each term). This is so because the natural variables 
of the dynamic cavity method are probabilities of spin histories of which there 
are exponentially many ($2^t$ for synchronous updates over time $[0,t]$ with time constant one).
Therefore, although the dynamic cavity equations themselves only involve a finite number of terms, summing them
nevertheless (in general) entails a number of operations which is exponentially large in $t$.
Fully asymmetric networks is a special case since the cavity equations can then be marginalized over
time with no loss of information, and the complexity ``in time'' disappears.
Alternative ``ways out'' investigated in the literature use additional assumptions on the
evolution law such as majority dynamics~\cite{Kanoria2011} (\textit{i.e.} linear dynamics with thresholding) 
or, more recently, unidirectional dynamics. Models of this latter type, where after a variable makes a transition 
from a state to another it can never go back, are represented by the zero-temperature random field Ising model (RFIM) \cite{ohta2010universal}, cascade processes \cite{altarelli2013large}, spread optimization problems \cite{altarelli2013optimizing} and several epidemic models as, for instance, the susceptible-infected-recovered (SIR) model \cite{lokhov2013inferring, lokhov2014dynamic}. The main peculiarity of such models is that the dynamics can be parametrized in terms of the time(s) at which the transition from one state of the variable to another occurs, which again eliminates time complexity. 

Another approach was taken in 
\cite{Neri2009}, where a simplifying ``one-time'' assumption was introduced,
which for fully asymmetric networks reduces to (exact) marginalization over time.
For the kinetic Ising model with synchronous and asynchronous updates this was
later shown to be considerably more accurate than dynamic mean-field, not only for asymmetric networks
but also for partly symmetric networks~\cite{aurell2011message,aurell2012dynamic}.
A different approach, based on variational approximations, was very recently proposed in \cite{pelizzola2013variational} 
where the author shows better performances in recovering stationary states compared to existing methods.
Unfortunately, except for fully asymmetric networks all these
approaches are limited to steady states, and hence cannot handle dynamic phenomena.

In this contribution we present a method to approximatively close the dynamic cavity equations
for synchronous updates with no assumptions on the underlying network and evolution law beyond that 
the network is locally tree-like and the evolution law is Markov. Unlike other methods already present in the 
literature, our approach is built not only to recover stationary states but potentially also the transient, i.e. the 
out of equilibrium dynamics. The method is built on two ingredients. The first, which already appeared in this context in~\cite{altarelli2013optimizing}, is the use of the graph expansion technique of~\cite{mezard2009information} 
to rewrite the probabilistic model in a way such that the underlying graph is explicitly
locally tree-like, and the standard cavity equations can be used.
The second ingredient is the assumption that a set of auxilary probability distributions on 
spin histories, ``messages'' in cavity method language, contain dependenciess that are mainly 
local in time. A closure of the dynamic cavity equations is then effectuated within the class of $n$-order Markov processes.
For definiteness we will here present the closure in the class of ordinary Markov processes ($n=1$)
and only outline the extensions to $n>1$, to which we intend to return in a future contribution.
The pioneering contribution~\cite{Neri2009} can, in present perspective, be seen as a
closure in the class of Bernoulli processes ($n=0$), without using graph expansion.

\section{The dynamic cavity equations}
\begin{figure}[] 
\begin{center}
\includegraphics[width= 5cm]{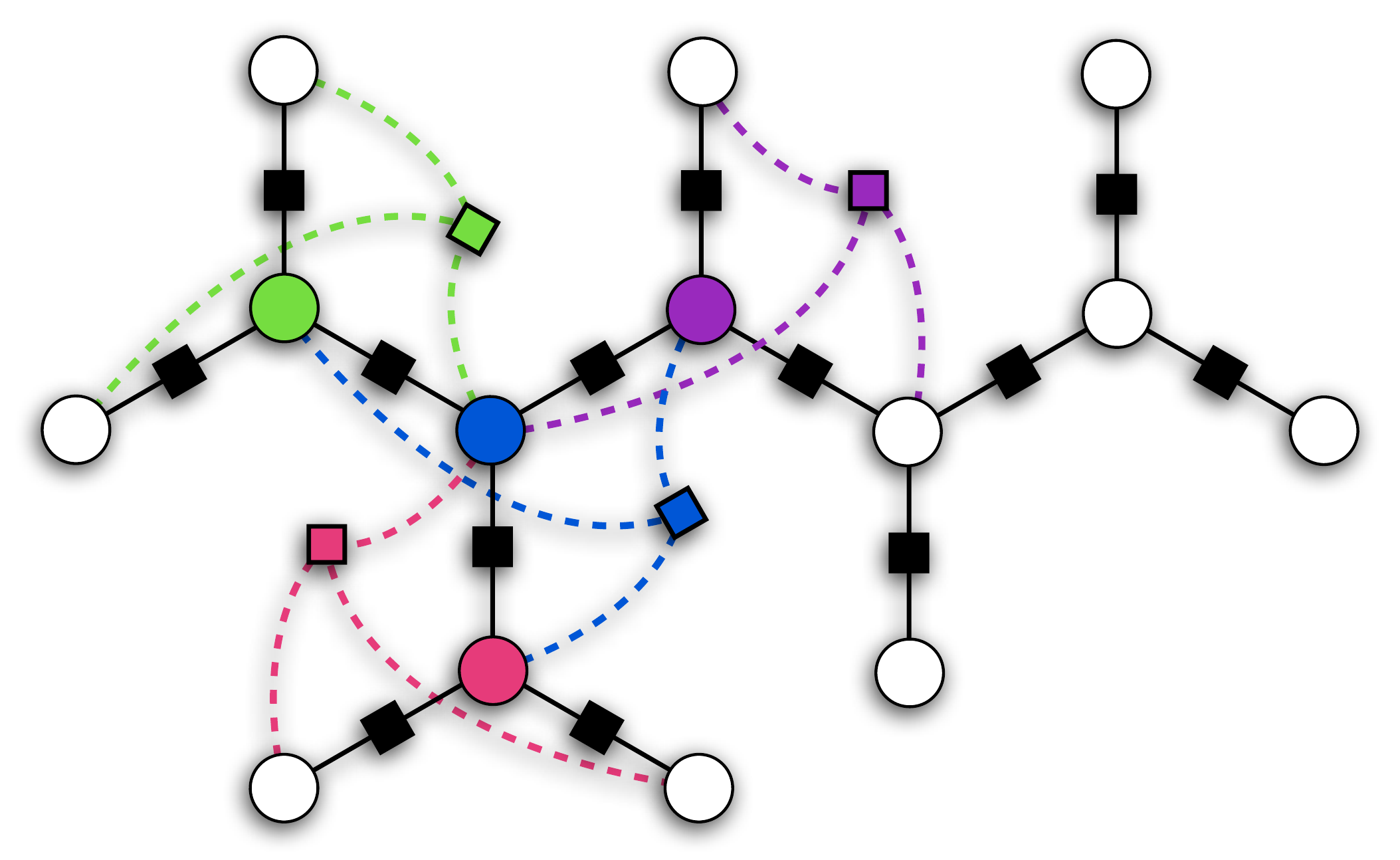}
\caption{Color online. Original graph topologically tree-like shown in a factor graph representation in which loops emerge naturally in time when dynamics is considered. Circles illustrate spin variables and squares interactions among them. More specifically, black squares picture the interaction between one spin and one of its neighbours, i.e. $\phi_{ij}(\s_i(t),\s_j({t-1}))$, whereas coloured squares picture the interaction among neighbours of a given spin, i.e. $\phi_j(\{\s_j(t-1)\}_{j \in \partial i})$.}
\label{fig:original_graph}
\end{center}
\end{figure}

\begin{figure}[] 
\begin{center}
\includegraphics[width= 4 cm, height= 3 cm]{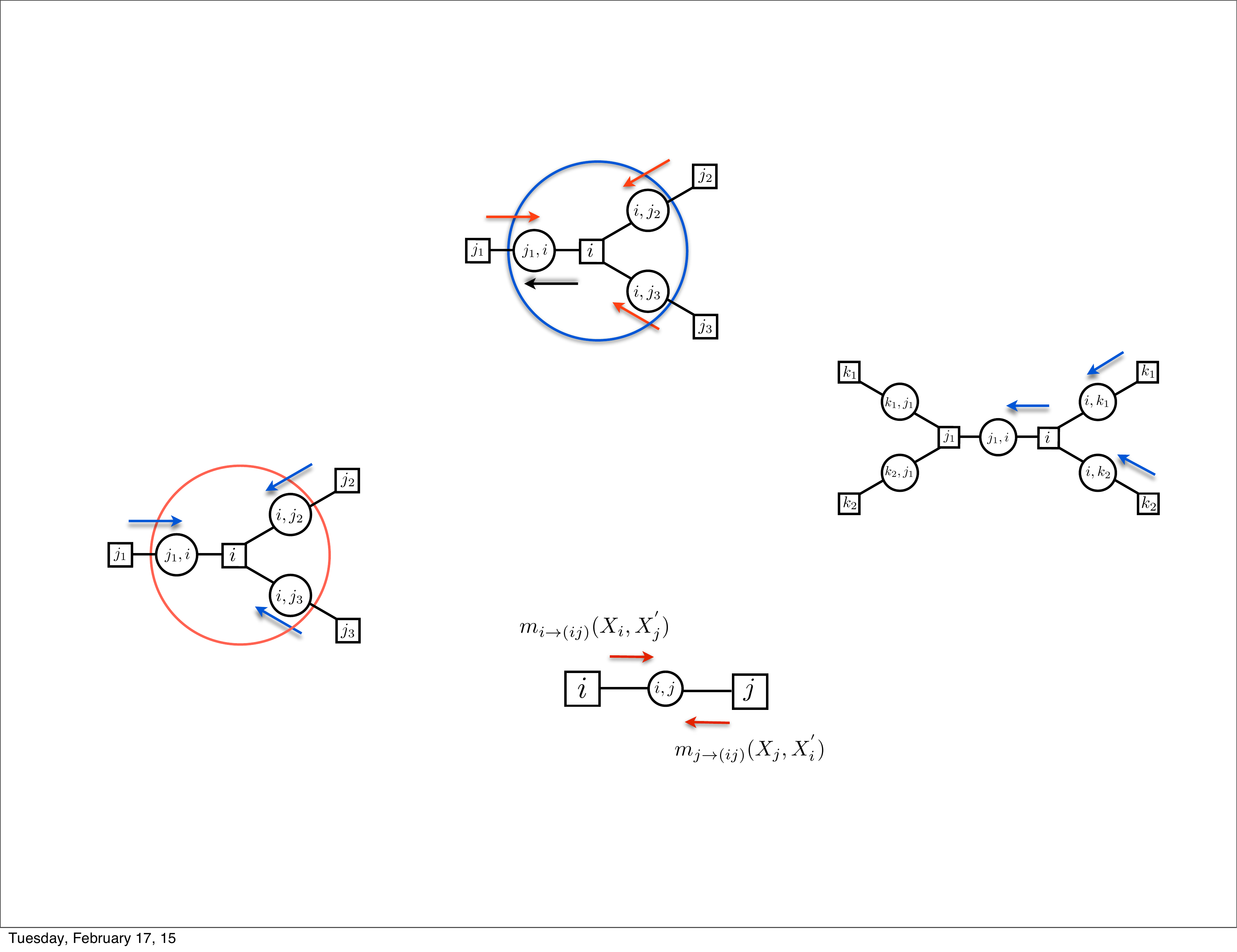}
\caption{Auxiliary graph obtained from the original one shown in Fig. \ref{fig:original_graph} where the time-loops 
have been removed with a standard procedure. The new variable nodes contain the spin history of two variables 
which were neighbours in the original graph and factor nodes contain the interactions needed to generate and pass this history, i.e. $\tilde{\Phi} = \delta_{X_i^{(ij_1)},\dots,X_i}  \prod_{s=1}^t w_i(\s_i^s | \s_i^{s-1}, \{\s_j^{s-1}\}_{j\in\partial i})P_0(\s_i,\{\s_j\}_{j \in \partial i}) $. Observe that the delta function is already summed over in the main text.}
\label{fig:aux_graph}
\end{center}
\end{figure}

We consider a probabilistic graphical model defined on a tree-like graph $\mathcal{G}=(V,E)$  where $V$ is a 
set of $N$ vertices and $E$ is a set of directed edges. Spins, Boolean variables $\s_i(t)= \pm 1$, are associated to each vertex at each 
time. We denote by $X_i= (\s_i(0),\s_i(1),\dots,\s_i(t))$ the spin history of spin $i$
and the evolution law $w_i(\s_i^s | \s_i^{s-1}, \{\s_j^{s-1}\}_{j\in\partial i})$ 
is the conditional probability of spin $i$ to take value $\s_i$ at time $s$ given 
the values of spins $i$ and $\partial i$, the graph neighbours of $i$, at time $s-1$.
We note that this class is larger than the previously investigated majority dynamics and (synchronous) kinetic Ising
models since we allow the evolution law to depend on $\s_i^{s-1}$.
The joint probability over the spin histories can be then written as
\begin{equation}
P(X_1,\dots,X_N) = \prod_{i \in V} \prod_{s=1}^t w_i(\s_i^s | \s_i^{s-1}, \{\s_j^{s-1}\}_{j\in\partial i})P_0
\label{eq:spin-history-probability}
\end{equation}
where $P_0=P(\s_1(0),\dots,\s_N(0))$ is the initial joint distribution at time zero and $t$ is the final time. 
We recall that the cavity method works well when the underlying network is (locally) tree-like. 
Being a conditional probability any evolution law $w_i$ can be written as $\exp\left(\s_i^s\Psi_i-\log 2\cosh\Psi_i\right)$
where $\Psi_i$ is some function of $(\s_i^{s-1}, \{\s_j^{s-1}\})$.  
$w_i$ hence contains two
types of ``interactions'', namely $\s_i^s\Psi_i$ and $\log 2\cosh\Psi_i$, and if
$P(X_1,\dots,X_N)$ in (\ref{eq:spin-history-probability})
is written as $\exp\left(F(X_1,\dots,X_N)\right)$ the graph describing the dependencies in $F$
will have short loops, which necessarly emerge  even if the network topology is tree-like, as illustrated in Fig \ref{fig:original_graph}. 
As shown in \cite{mezard2009information,altarelli2013optimizing} such loops can 
however be removed by defining an auxiliary factor graph at the price that the new variable nodes will contain more that one old variable. 
The procedure consists in changing the old variable nodes into factor nodes and changing the old interaction edges into new variable nodes.
The resulting topology of the new expanded graph is shown in Fig. \ref{fig:aux_graph}
where each variable node now contains the spin histories of two spins. The functions sitting on the new factor nodes are 
defined in order to guarantee all interactions to remain the same as in the original graph and so, for consistency, the spin variables of the same 
type that appear on different (new) variable nodes have to take the same value. 
In the auxiliary graph the node $(i,j)$ contains the pair $(X_i^{(ij)}, X_{j}^{(ij)})$ where by $X_i^{(ij)}$ 
we mean a variable of the same kind as spin history $X_i$ residing in the node $(i,j)$. The joint probability 
distribution of the new variables on the expanded factor graph of Fig. \ref{fig:aux_graph} can then be written as:
\begin{align}
P(\{ X_i^{(ij)}, X_j^{(ij)}\}_{(ij) \in E}) =& \prod_{i \in V} \delta_{X_i^{(ij_1)},\, X_i^{(ij_2)},\dots,X_i}  \prod_{s=1}^t w_i(\s_i^s | \s_i^{s-1}, \{\s_j^{s-1}\}_{j\in\partial i})P_0
\end{align}
where all the variables in $w_i(\s_i^s | \s_i^{s-1}, \{\s_j^{s-1}\}_{j\in\partial i})$ are taken from the surrounding spin histories 
$(X_i^{(ij_1)}, X_{i}^{(ij_2)}, \dots)$ and the constraint enforces that these histories agree. If the original graph is (locally) 
tree like, this procedure removes the loops in time and gives us a new auxiliary graph which is also (locally) tree like. 
The standard belief propagation (BP) update equations can then be written, for the variables in the new graph, as in the static case (see eq. (14.15) in \cite{mezard2009information}) as:
\begin{align} \label{BPeq} 
m_{i \to (ij)} (X_i^{}, X_j^{}) \propto &\sum_{\{ X_k^{} \}}  \Phi(X_i^{}, X_j^{},\{X_k^{}\}) \prod_{k \in \partial i \backslash j} m_{k \to (ik)} (X_k^{}, X_i^{})
\end{align}
where $ \Phi(X_i^{}, X_j^{},\{X_k^{}\})=P_0 \prod_{s=1}^{t} w_i (\s_i^{s} | \s_i^{s-1}, \{\s_k^{s-1}\}_{k \in \partial i \backslash j}, \s_j^{s-1})$
is analogous to a potential sitting in the factor nodes which transfers the dynamics to its neighbour variable nodes and the normalization factor can be computed by the condition that $\sum_{X_i, X_j} m_{i \to (ij)} (X_i^{}, X_j^{})=1$. Above we have shortened the notation to $X_i=X_i^{(ij)}$, $X_k=X_k^{(ik)}$ and the same for $X_j$ and all the $\s$-variables contained in $w_i$. Let us note that we have fewer distinct messages respect to the static BP formulation since messages from factor to variable nodes are, in this case, the same as messages from variable to factor nodes, i.e. $m_{j \to (ij)} (X_j,X_i) = m_{(ij) \to i} (X_j,X_i)$. Hence, for the topology shown in Fig \ref{fig:aux_graph}, the message on the LHS of \eqref{BPeq} is illustrated in black whereas the messages on the RHS are those two in red coming from the right side of the picture.  To simplify notation, from now on variables with no apex refer to variables at time $t$, i.e. $\s_i=\s_i(t)$, 
whereas variables with apex(es) refer respectively to previous time(s), i.e. $\s_i'=\s_i(t-1), \s_i''=\s_i(t-2),\dots$ and so $X_i^{'}$ is the spin history of spin $i$ up to time $t-1$.
Let us observe that, while notationally compact, equations (\ref{BPeq}) is, except for short times, 
computationally intractable as the right-hand side involves sums over complete spin histories. 

\section{Markovian closure of the dynamic cavity equations}

We first observe that, given the Markovianity of the dynamics contained in $\Phi(X_i^{}, X_j^{},\{X_k^{}\})$, the messages 
$m_{i \to (ij)} (X_i^{}, X_j^{})$ in (\ref{BPeq})  
actually do not depend on the spin variable $\s_j$ at time $t$ but only at time $t-1$ and earlier times.
The messages $m_{i \to (ij)} (X_i^{}, X_j^{})$ are hence
always uniform distributions on $\s_j(t)$ and, since $\s_j(t)$ is therefore a kind of dummy argument,
we may simplify the notation by writing $m_{i \to (ij)} (X_i^{}, X_j^{'})$.
Furthermore, if we use the same simplification on the right hand side of
(\ref{BPeq}) we have incoming messages $m_{k \to (ik)} (X_k^{}, X_i^{'})$
which can be marginalized to $m_{k \to (ik)} (X_k^{'}, X_i^{'})$ as there is
no other dependence on $\s_k$, the value of spin $k$ at time $t$. Using Markovianity 
again we can simplify further to $m_{k \to (ik)} (X_k^{'}, X_i^{''})$ and write (\ref{BPeq}) as 
\begin{align} \label{BPeq-2} 
m_{i \to (ij)} (X_i^{}, X_j^{'}) \propto &\sum_{\{ X_k^{'} \}}  \Phi(X_i^{}, X_j^{'},\{X_k^{'}\}) \prod_{k \in \partial i \backslash j} m_{k \to (ik)} (X_k^{'}, X_i^{''})
\end{align}
Equation (\ref{BPeq-2}) shows that the dynamic cavity equations have a different structure 
and are actually, in some respects, simpler than standard BP updates, since if all messages up to some
time $t-1$ are known, messages up to time $t$ can be evaluated directly without the need
to iterate to a fixed point. 
\noindent
\subsection{Fully asymmetric graph}
It is worth to observe how equation (\ref{BPeq-2}) simplifies when referred to a fully asymmetric graph, with interaction couplings between node $i$ and $j$ such that $J_{ij} \neq 0$ and $J_{ji} =0$. Under this assumption the interaction function $\Phi$ no longer depends on the spin history $X_j^{'}$ and, as a consequence, the message on the LHS of the equation does not depend on that history either, since such dependence is carried in only through the function $\Phi$. For consistency, we can apply the same argument to the messages on the RHS and conclude that they only depend on the spin history $X_k^{'}$ and no longer on $X_i{''}$. Then remembering that $\Phi$ is a normalized function respect to the variable $\s_i^s$, as shown in the text below equation \eqref{BPeq}, we can sum both sides of \eqref{BPeq-2} over the spin history $X_i^{'}$ and then make use of the sum over $\{X_k^{'}\}$ on the RHS to obtain the simplified version of the dynamic message-passing equation for fully asymmetric graph
\begin{align} \label{DMPfullyAsy} 
m_{i \to (ij)} (\s_i^{t}) = &\sum_{\{ \s_k^{t-1} \}_{k \in \partial i \backslash j}}  w_i(\s_i^{t}| \{\s_k^{t-1}\}_{k \in \partial i \backslash j}) \prod_{k \in \partial i \backslash j} m_{k \to (ik)} (\s_k^{t-1}).
\end{align}
This equation is agreement with previous literature for the same graph topology \cite{Neri2009,aurell2011message,aurell2011three}.
\subsection{Graph with arbitrary connectivity symmetry}
In what follows we propose a Markovian closure of the dynamic $\emph{BP update}$ equation \eqref{BPeq-2} for a network with arbitrary connectivity symmetry. In the next section the same closure is used to derivate the dynamic \emph{BP output} equations. In a closure in the class of $n$-th order Markov processes we assume
$m_{i \to (ij)} (X_i^{}, X_j^{'}) = \prod_{s=n}^t  T^{(n)}_{i \to (ij)} (\s_i^{s} | \s_i^{s-1}, \s_j^{s-1},\ldots,\s_i^{s-n}, \s_j^{s-n})$
and solve (\ref{BPeq-2}) iteratively. We here consider $n=1$. The
marginalizations over the last and last two times of the variable node $(i,j)$ are
\begin{align}\label{P1time}
P_{i \to (ij)}^{(t-1)}(\s_i^{'},\s_j^{'}) &= \sum_{\substack{X_i^{''}, X_j^{''}}} m_{i \to (ij)} (X_i^{'}, X_j^{''})\\ \label{P2times}
P_{i \to (ij)}^{(t,\, t-1)}(\s_i^{}, \s_i^{'}, \s_j^{'}) &= \sum_{X_i^{''},\, X_j^{''}} m_{i \to (ij)} (X_i^{}, X_j^{'}), 
\end{align}
by assumption linked by 
\begin{equation}
P^{(t,\, t-1)}(\s_i, \s_i^{'}, \s_j^{'}) = T(\s_i | \s_i^{'}, \s_j^{'}) \, P^{(t-1)}(\s_i^{'},\s_j^{'})
\label{eq:T}
\end{equation}
where we omitted the subscript $i \to (ij)$ and the superscript time dependence of $T^{(t,t-1)}$ for readability.
We note that by above $P^{(t-1)}(\s_i^{'},\s_j^{'})$ actually does not depend on its second argument
and we will therefore from now on simplify to $P^{(t-1)}(\s_i^{'})$.
Closure means to make the same assumptions for the upstream messages
$m_{k \to (ik)} (X_k^{}, X_i^{'})$, use (\ref{BPeq-2}) to compute 
$P_{i \to (ij)}^{(t-1)}$ and $P_{i \to (ij)}^{(t,\, t-1)}$ in
(\ref{P1time}) and (\ref{P2times}), and then take
(\ref{eq:T}) to define $T(\s_i | \s_i^{'}, \s_j^{'})$.
This can be done by introducing an auxiliary function  $\mathcal{F}$
\begin{align} \label{Fdef} 
\mathcal{F}_{i \to (ij)}^{(t-1)} (\s_i^{'},\{\s_k^{'}\}_{k \in \partial i \backslash j}) &=\hspace{-0.5cm} \sum_{\substack{\{ X_k^{''},\, X_i^{''}, \, X_j^{''}\}}} \prod_{k \in \partial i \backslash j} \, m_{k \to (ik)} (X_k^{'}, X_i^{''}) \prod_{s=1}^{t-1} w_i (\s_i^{s} | \s_i^{s-1}, \{\s_k^{s-1}\}_{k \in \partial i \backslash j}, \s_j^{s-1})P_0
\end{align}
which is strictly a specific marginalization of the right hand side of (\ref{BPeq-2}).
From now on, we use the notation $\{\s_k \}_{\backslash}=\{ \s_k \}_{k \in \partial i \backslash j}$ and $\prod_{k\backslash}=\prod_{k \in \partial i \backslash j}$. In terms of \eqref{Fdef} 
we have
\begin{align}\label{BP1}
P_{i \to (ij)}^{(t-1)}(\s_i^{'}) &\propto \sum_{\{\s_k^{'}\}}\mathcal{F}_{i \to (ij)}^{(t-1)} (\s_i^{'},\{\s_k^{'}\}_{\backslash})\\
P_{i \to (ij)}^{(t,\, t-1)}( \s_i^{}, \s_i^{'}, \s_j^{'})&\propto \sum_{\{\s_k^{'}\}} \, w_i (\s_i^{} | \s_i^{'}, \{\s_k^{'}\}_{\backslash}, \s_j^{'}) \, \mathcal{F}_{i \to (ij)}^{(t-1)} (\s_i^{'},\{\s_k^{'}\}_{\backslash}) \label{BP2}
\end{align}
On the other hand, by the (assumed) Markovianity of the upstream messages we can write an iterative equation 
in time for $\mathcal{F}$:
\begin{align}\label{eq:iterF}
\mathcal{F}_{i \to (ij)}^{(t)} (\s_i^{},\{\s_k^{}\}_{\backslash}) &=\sum_{\{\s_k^{'}\}, \s_i^{'}, \s_j^{'}} \prod_{k \backslash} T_{k \to (ik)} (\s_k^{} | \s_k^{'}, \s_i^{'}) \, w_i (\s_i^{} | \s_i^{'}, \{\s_k^{'}\}_{\backslash}, \s_j^{'})\,\mathcal{F}_{i \to (ij)}^{(t-1)}(\s_i^{'},\{\s_k^{'}\}_{\backslash}).
\end{align}
Whereas the iterative equation for $T$, using (\ref{BP1}) and (\ref{BP2}), reads as
\begin{align} \label{eqT} 
T_{i \to (ij)}^{} (\s_i^{} | \s_i^{'}, \s_j^{'}) = \frac{\sum_{\{\s_k^{'}\}} w_i (\s_i^{} | \s_i^{' }, \{\s_k^{'}\}_{\backslash}, \s_j^{'}) \, \mathcal{F}_{i \to (ij)}^{(t-1)} (\s_i^{'},\{\s_k^{'}\}_{\backslash})}{\sum_{\s_i^{},\, \{\s_k^{'}\}} w_i (\s_i^{} | \s_i^{'}, \{\s_k^{'}\}_{\backslash}, \s_j^{'}) \, \mathcal{F}_{i \to (ij)}^{(t-1)} (\s_i^{'},\{\s_k^{'}\}_{\backslash}) },   
\end{align}
where the denominator takes care of the normalization. Equation \eqref{eq:iterF} and \eqref{eqT}, which solve the dynamic cavity equations under the assumption that messages are Markovian, are the first result of this paper and represent the ingredients to solve the BP update equations at any time.
The entire procedure clearly takes polynomial time instead of an exponential time as the original formulation
and, as noted above, does not involve iteration to fixed point.

\section{Markovian closure of the BP output equations}
We now turn to the \emph{BP output equations}, i.e. the equations for the actual marginal probability both on the auxiliary and on the original factor graph. The marginal probability of one variable node $(i,j)$ in the auxiliary graph is simply given by the product of the incoming messages to the node $(i,j)$:
\be \label{ProbAux} 
P_{(ij)}(X_i^{}, X_j^{})\propto m_{i \to (ij)}(X_i^{}, X_j^{'}) \, m_{j \to (ij)}(X_j^{},X_i^{'}).
\ee
We are however interested in the single-site one-time marginal probability on the original graph, $P_i(\s_i(t))$, from which we can compute physical observables of interest such as, for instance, the magnetization at any given time. We can get this marginal starting from the one-site marginal probability on a spin history, $P_i(X_i)$. 
In terms of the auxiliary probability distribution on the expanded graph this reads as 
\begin{align}\label{Porig} 
P_i(X_i) &\propto \sum_{X_j^{}}  \prod_{j \in \partial i} m_{j \to (ij)}(X_j, X_i^{'}) \Big[ \prod_{s=1}^t w_i(\s_i^s | \s_i^{s-1},\{ \s_j^{s-1}\}_{j \in \partial i}) P_0\Big],
\end{align}
where spins $j$'s are the neighbours of site $i$ in the original and expanded graph (see Fig \ref{fig:aux_graph}).
To solve these equations on the same level of approximation as the update equation 
we define a new auxiliary function
\begin{align} 
\mathcal{G}^{(t-1)}(\s_i^{'},\{ \s_j^{'}\}_{j \in \partial i}) =& \sum_{X_{i,}^{''} \{X_j^{''}\}_{j \in \partial i}}  \prod_{j \in \partial i} m_{j \to (ij)}(X_j^{'}, X_i^{''}) \Big[\prod_{s=1}^{t-1} w_i(\s_i^{s} | \s_i^{s-1},\{ \s_j^{s-1}\}_{j \in \partial i}) P_0 \Big].
\end{align}
The single-site one-time and two-time marginal probabilities on the original graph then follow from \eqref{Porig}: 
\begin{align} \label{origP1}
P^{(t-1)}_i(\s_i^{'}) &\propto \sum_{ \{ \s_j^{'}\}_{j \in \partial i}}\mathcal{G}^{(t-1)}(\s_i^{'},\{ \s_j^{'}\}_{j \in \partial i}),\\ 
P^{(t, \, t-1)}_i(\s_i, \s_i^{'}) &\propto \sum_{ \{ \s_j^{'}\}_{j \in \partial i}} w_i(\s_i | \s_i^{'},\{ \s_j^{'}\}_{j \in \partial i})\, \mathcal{G}^{(t-1)}(\s_i^{'},\{ \s_j^{'}\}_{j \in \partial i}). \label{origP2}
\end{align}
In analogy to $\mathcal{F}$ above, we can write a recursive equation for $\mathcal{G}$ by using the 
Markovian assumption for messages:
\begin{align}\label{eq:iterG} 
\mathcal{G}^{(t)}(\s_i,\{ \s_j^{}\}_{j \in \partial i}) &= \sum_{ \s_i^{'}, \{\s_j^{'}\}_{j \in \partial i}} \prod_{j\in \partial i}  T_{j \to (ij)} (\s_j^{} | \s_j^{'}, \s_i^{'}) \, w_i(\s_i | \s_i^{'},\{ \s_j^{'}\}_{j \in \partial i}) \, \mathcal{G}^{(t-1)}(\s_i^{'},\{ \s_j^{'}\}_{j \in \partial i})
\end{align}
Hence starting with initial value for the functions $T$, $\mathcal{F}$ and $\mathcal{G}$ equations \eqref{eq:iterF},  \eqref{eqT} and \eqref{eq:iterG} can be iterated up to the desired time and equation \eqref{origP1} can be used to compute the time-dependent marginal probability of site $i$. We highlight that unlike the original formulation, which takes an exponential time to compute marginals, the entire scheme presented here has a polynomial computational cost.

\section{Results}

In this section we test the accuracy of our dynamic message-passing (DMP) approach on a statistical physics model often chosen as a case of study to investigate dynamics of complex systems: the kinetic Ising model. 

\subsection{A case of study: the kinetic Ising model with arbitrary connectivity symmetry}

We compare the performance of DMP to Monte Carlo Markov Chain simulations (MC) with Glauber dynamics on the kinetic Ising model on a random diluted graph with arbitrary connectivity symmetry: fully symmetric, partially asymmetric and fully asymmetric network. The comparison is performed computing the behaviour of the magnetization of the model both during the transient and at equilibrium (or at the stationary states when detailed balance does not hold). Following \cite{hatchett2004parallel} we introduce a connectivity matrix $c_{ij}$ where $c_{ij} = 1$ if there is a link from vertex $i$ to vertex $j$, $c_{ij}$ = 0 otherwise, and matrix elements $c_{ij}$ and $c_{kl}$ are independent unless $\{kl\} = \{ji\}$. The following distributions then specify the graph topology: the marginal one-link distribution
\be
p(c_{ij})= \frac{c}{N} \delta_{1,c_{ij}} + (1-\frac{c}{N})\delta_{0,c_{ij}}
\ee
and the conditional distribution
\be
p(c_{ij}|c_{ji})= \epsilon \delta_{c_{ij}, c_{ji}} + (1-\epsilon)p(c_{ij}).
\ee
Above, $N$ is the size of the network, $c$ the average connectivity and $\epsilon \in [0,1]$ a parameter which controls the asymmetry. The value $\epsilon= 0$ gives a fully asymmetric network whereas $\epsilon=1$ a fully symmetric one.

\begin{figure}[t!] 
\begin{center}
\includegraphics[width= 17cm, height= 4.cm]{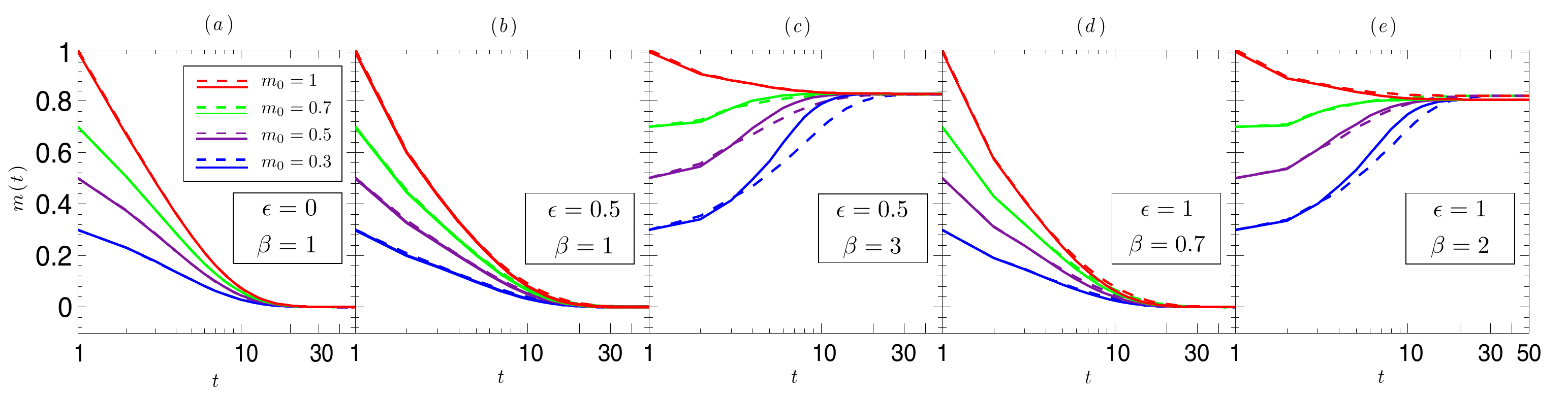}
\caption{Color online. Comparison between dynamic message passing algorithm (solid lines) and Monte Carlo simulation (dashed lines) for the evolution of the magnetization $m(t)$ in an Erd\"os R\'enyi network of $N = 5000$ nodes and average connectivity $c=3$. Different colors represent different initial magnetizations $m(0)$, from the largest (upper line) to the smallest (lower line) according to the legend. (b) and (c) plot: partially asymmetric network ($\epsilon = 0.5$) respectively at low and high temperatures.  (d) and (e) plot: fully symmetric network respectively at low and high temperatures. In MC simulations $m(t)$ is averaged over 5000 samples.}
\label{fig:results}
\end{center}
\end{figure}

\begin{figure}[t!] 
\begin{center}
\includegraphics[width= 17cm, height= 4.cm]{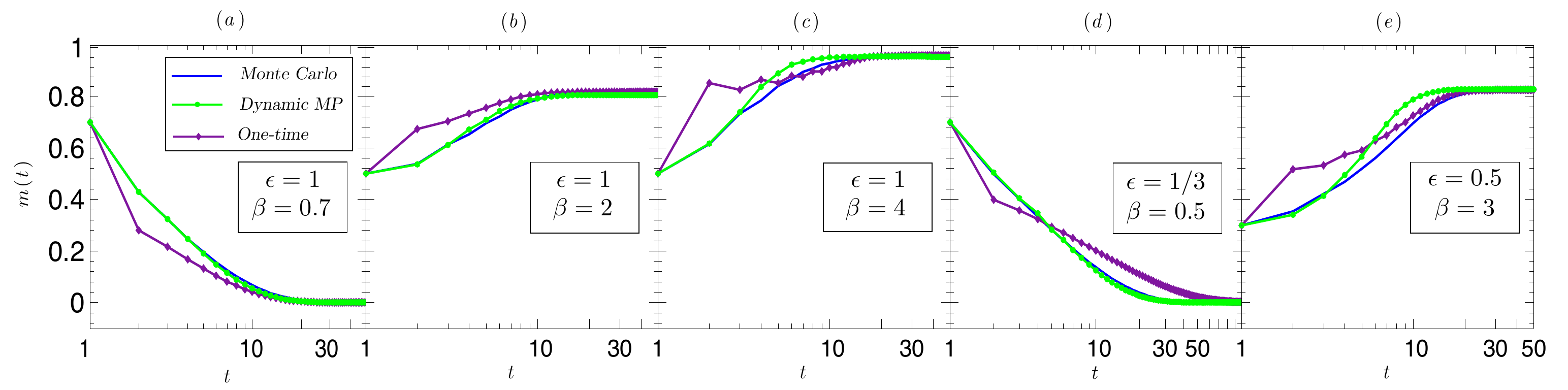}
\caption{Color online. Comparison between dynamic message passing algorithm (solid lines with $\bullet$ dots), one-time approximation (solid lines with $\scriptstyle\blacklozenge$ dots) and Monte Carlo simulation (solid lines) for the evolution of the magnetization $m(t)$ in an in an Erd\"os R\'enyi network of $N = 5000$ nodes and average connectivity $c=3$. (a), (b) and (c)-plot: fully symmetric network at high and low temperatures. (d) and (e)-plot: partially symmetric networks with different connectivity symmetry at high and low temperatures. In MC simulations $m(t)$ is averaged over 5000 samples.}
\label{fig:comparison}
\end{center}
\end{figure}

In the kinetic Ising model the transition probability rate $w_i$, which appears above, is given by $w_i(\s_i^s | \{\s_j^{s-1}\})= \exp{(\beta \s_i^s J_{ij} \s_j^{s-1} - \log 2 \cosh(\beta \sum_j J_{ij} \, \s_j^{s-1}}))$ where the $j$'s variable are neighbours of spin $i$, $J_{ij}$ are the interaction strengths between site $i$ and $j$ and $\beta$ is the inverse temperature. We observe that this model belongs to a subclass of  the models considered within the general formulation above as the transition rate does not depend explicitly on $\s_i^{s-1}$. Since we here want to mainly investigate the effect of the asymmetry on the performances of our algorithm,  we restrict our numerical analysis to ferromagnetic models, i.e. interaction strength $J_{ij} = J >0$ for every pair of sites $i,j$. For simplicity, all the spins on the graph are chosen independent at the initial time although correlated initial conditions could also be considered in the above formulation. Results are shown in Figure \ref{fig:results} for several values of the initial magnetization at different temperatures and for various values of the asymmetry parameter $\epsilon$.
Numerical results for fully asymmetric networks are obtained by using the simplified version \eqref{DMPfullyAsy} of the dynamic message-passing equation and, as expected from  \cite{Neri2009,aurell2011message,aurell2012dynamic}, they show a perfect agreement with Monte Carlo simulations both for the computation of the transient and the stationary state (see panel $(a)$). When the full asymmetry is broken and a network with feedback is considered numerics is obtained through the DMP scheme presented above. Comparison with MC shows that the results provided by DMP are still very good for high enough temperature (above the critical transition) both for partially and fully symmetric connectivities. Indeed as it is possible to note from Fig \ref{fig:results} (partially asymmetric network in panel $(b)$ and fully symmetric one in panel $(d)$) the transient regime is very well reproduced by DMP and the stationary (or equilibrium) state is the same as MC. When temperature is decreased below the critical ferromagnetic transition the performances of DMP start getting worse (panel $(c)$ and $(e)$). In this regime, the transient is very well recovered only for the first few initial steps of the dynamics and gets progressively worse for longer times. Nevertheless, for the partially asymmetric network considered $(\epsilon= 0.5)$ the stationary state reached by DMP and MC simulations coincide, although DMP has a faster convergence to that (panel $(c)$). For fully symmetric networks similar considerations follow for the transient regime although the equilibrium state reached by DMP and MC simulations at this temperature are slightly different (panel $(e)$). This is not always the case indeed we noticed numerically that, for some of the temperatures even below the critical transition, DMP reaches the same equilibrium state as MC (see for instance Fig \ref{fig:comparison}, panel (c)). However, for fully symmetric networks, it is known that the stationary solutions of the one-time approximation (OTA) method presented in \cite{Neri2009,aurell2011three} are also solution of the static BP equations. Therefore the agreement  between MC simulations and the one-time approximation for the equilibrium state of fully symmetric networks is expected to be near perfect whereas, for some temperatures, the DMP approach proposed here presents small differences with the equilibrium MC solution. 

In order to investigate the performances of DMP respect to the one-time approximation, we compared both algorithms with MC simulations for different temperatures and for various values of the asymmetry parameter $\epsilon$. The numerical investigation, partially illustrated in Fig \ref{fig:comparison}, shows that regardless the network connectivity symmetry DMP always outperforms OTA for the transient dynamic regime. For temperatures above the critical transition ($T_c$) both algorithms converge to the same stationary or equilibrium state in agreement with MC simulations (see panel (a) and (d)) and, surprisingly, also for partially asymmetric networks the one-time approximation recovers well the MC stationary solution (panel (d)).
For temperatures lower that $T_c$ the following picture emerges. For fully symmetric networks OTA always reaches the same equilibrium state as MC whereas DMP in some cases does (see panel (c)) in some cases does not (see panel (b)), depending on the temperature. For partially asymmetric networks the performances of DMP get better and, surprisingly, also OTA gives pretty good results for the stationary states (see panel (e)). The differences in the reconstruction of the transient and stationary regime of the DMP and the one-time approach can be understood remembering that the two algorithms do not belong to the same class of approximations.  \\
We conclude this section underling that we expect improvement of the results provided by the dynamic message-passing scheme presented, both for the transient and for the stationary or equilibrium state, when a higher order closure of the DMP equations is made (see next section). 
\section{Higher order closures}
To close (\ref{BPeq}) in the class of $n$-th order Markov processes we consider,
instead of $P_{i \to (ij)}^{(t-1)}$ and $P_{i \to (ij)}^{(t,t-1)}$ in
(\ref{BP1}) and (\ref{BP2}), the marginalizations 
$P_{i \to (ij)}^{(t-1,t-2,\ldots,t-n)}$ and 
$P_{i \to (ij)}^{(t,t-1,t-2,\ldots,t-n)}$ and auxiliary functions $F^{(n)}$ and $G^{(n)}$ which
depend on $n$ earlier times.
The time iteration of $F^{(n)}$, $G^{(n)}$ and the iterative solution of the 
$n$-th order kernel $T^{(n)}_{i \to (ij)} (\s_i^{} | \s_i^{'}, \s_j^{'},\ldots,\s_i(t-n), \s_j(t-n))$
then proceed analogously to above.
The computational cost obviously increases quickly with $n$ and the value of higher order closures
will depend on the application and the model. This issue will be addressed in future contributions.

\section{Acknowledgements}

The authors acknowledge valuable discussions with 
Federico Ricci-Tersenghi, Silvio Franz,
Haiping Huang, Andrey Y. Lokhov, Yoshiyuki Kabashima and Gabriele Perugini.  
This work was finalized during the program "Collective Dynamics in Information System" at Kavli Institute for Theoretical Physics China (KITPC) in Beijing, China. It has been funded under FP7/2007-2013/grant agreement n¡ 290038 (GDF) and supported by the Swedish Science Council through grant 621-2012-2982 and by the Academy of Finland through its Center of Excellence COIN (EA).

\bibliographystyle{unsrt}
\bibliography{general_2}
\end{document}